\documentstyle[epsf]{article}

\newcommand{\be}{\begin{equation}}
\newcommand{\ee}{\end{equation}}

\begin{document}

\begin{flushright}
Liverpool Preprint: LTH 514\\
 \end{flushright}
  
\vspace{5mm}
\begin{center}
{\LARGE \bf The flavour singlet mesons in QCD.}\\[10mm] 
{\large\it UKQCD Collaboration}\\[3mm]
 
 {\bf   C. McNeile,   C.~Michael and K. J. Sharkey\\ 
Theoretical Physics Division, Dept. of Mathematical Sciences, 
          University of Liverpool, Liverpool L69 3BX, UK }\\[2mm]

\end{center}

\begin{abstract}

 We study the flavour singlet mesons  from first principles using
lattice QCD. We explore the splitting between  flavour  singlet and
non-singlet for vector and axial  mesons as well as the more commonly
studied cases of the scalar and pseudoscalar mesons.

\end{abstract}
%


\section{Introduction}

 The splitting in mass of flavour singlet and non-singlet mesons  with
the same quark content arises from gluonic interactions. The  assumption
that these are small is known as the OZI rule. For the pseudoscalar
mesons  this splitting is not small (it is related to the $\eta$,\
$\eta'$ mass difference), basically because of the impact of the
anomaly. For scalar mesons the splitting is also expected to be large
because of  mixing with the nearby scalar glueball. It is usually
assumed that the  OZI rule is in good shape for the vector and axial
mesons. This is difficult to check experimentally because of mass shifts 
induced by decays etc. 

 Here we study this splitting from first principles using lattice QCD. 
We work in a simplified world with $N_f=2$ flavours of degenerate
quarks.  This is still sufficient to explore the sign and magnitude of
any splitting which can then have a phenomenological impact on the 
interpretation of the observed spectrum. In this $N_f=2$ world, isospin
is exact  so we classify the flavour singlet state as $m_0$ and the
non-singlet as $m_1$. We indeed do find that the splitting  is largest
for the scalar and pseudoscalar mesons, but we are able to  estimate the
magnitude of the splitting for vector and axial (both $f_1,\ a_1$ and
$h_1,\ b_1$  types of axial) mesons. For the vector and axial channels,
the corresponding glueball masses are known from lattice
studies~\cite{ukqcdgb,mp} to be  much heavier and so the expectation is
that the splitting $m_0-m_1$ due to such glueball effects  would be
small and negative (since mixed states repel). This is the same sign as
found for scalar mesons but opposite to that  for pseudoscalar mesons.
The usual explanation  for  the splitting in the pseudoscalar channel is
through the   topological charge density in the vacuum~\cite{inst}.

In lattice studies it is possible to measure separately  the 
non-singlet contribution which is given by connected correlation $C(t)$
while  the flavour singlet contribution has an additional disconnected
correlation  $D(t)$. 
 Previous lattice studies have been made of the  pseudoscalar
mesons~\cite{kuramashi,cppacs,MQA,sesam,cmcmp} and  scalar mesons
~\cite{lw,cmcm}. For a discussion including some results for vector and
axial mesons, see~\cite{isgur}.

At large t where ground state contributions dominate  we have 
 \be  C(t) =  c e^{-m_1 t}  \ee
 and 
 \be  C(t)+D(t) = d e^{-m_0 t}  \ee
 where  $m_0$ is the flavour singlet mass  and $m_1$ the flavour
non-singlet mass. Now if the same meson creation and destruction
operators are used for the  study of  both correlations, with quarks
degenerate in   mass,  $d$ and $c$ have the same sign. 

 Then by a study of  $D/C$ which is given  by 
 \be D/C= (d/c) e^{(m_1-m_0) t} -1 \label{dbyc.eq} \ee
 one can explore the mass splitting between flavour singlet and
non-singlet. We illustrate this behaviour in fig.~1, assuming that  only
ground state contributions occur at all $t$ values. Although it might be
 thought that $d=c$, we have shown previously~\cite{cmcm} that this is 
not necessarily the case, and indeed sign changes in $D/C$ versus $t$ 
can be required.
  So, in summary,  the slope (increase/decrease) of $D/C$ on a lattice
can determine the  sign and magnitude of $m_1-m_0$.

 This has already been explored in detail by us for the cases of 
pseudoscalar mesons~\cite{cmcmp} (where $m_0-m_1 > 0$ ) and  scalar
mesons~\cite{cmcm} (where $m_0-m_1 < 0$). Here we extend the study to 
vector and axial mesons.

\begin{figure}[th]

\epsfxsize=11cm\epsfbox{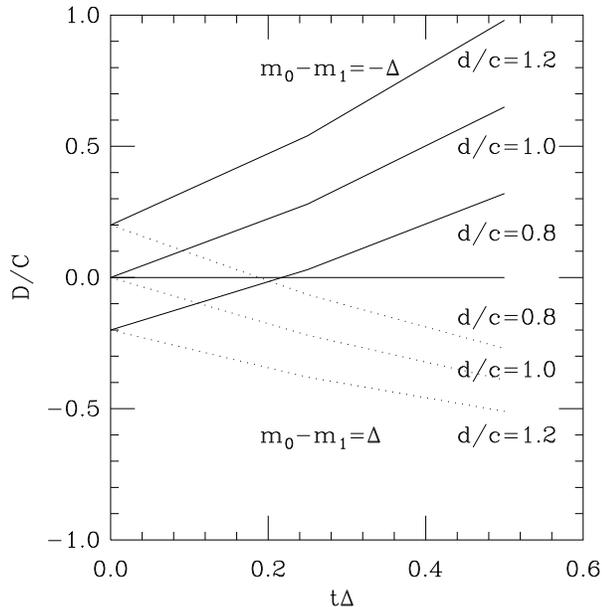}

 \caption{ The ratio of disconnected to connected contributions 
as given by eq.~\ref{dbyc.eq}.
 }
 \label{dbyc.fg}

\end{figure}

\section{Connected and Disconnected Contributions}

 Here we discuss in general what information is available  on the
connected correlation $C$ and disconnected correlation $D$ and their
relative sign.

For this section we consider only local operators to create a meson, 
namely $\bar{\psi}_A \Gamma \psi_B$, where subscripts refer to the 
flavour of the quarks. We choose a  basis of hermitian matrices for 
$\Gamma$, see Table~1.
 Here we restrict ourselves to spatially symmetric operators, a fuller
description which takes into account also the lattice cubic  symmetry is
given in~\cite{hybprd}.

 For $t > 1$ we can use reflection positivity to determine the sign of
$C$. 
 This corresponds to using a meson sink  $\bar{\psi}_B \Gamma^R
\psi_A$ with   $\Gamma^R= \gamma_4 \Gamma^\dag \gamma_4 = r_4 \Gamma$. 
 Then the correlation
 \be C(t) =(\bar{\psi}_A \Gamma   \psi_B)_0 
           (\bar{\psi}_B \Gamma^R \psi_A)_t  \ee
 will be positive for $t > 1$. 
 We evaluate this correlation in the usual way for the connected
correlation,  combining the Grassmannian fermion fields into propagators
 which yields another minus sign, and using the $\gamma_5$ hermitivity
of the  Wilson-Dirac fermion matrix to relate these propagators to a
common source. We indeed find that $C(t) > 0$ in our  measured
correlations for $t \ge 1$.

 In the flavour singlet case, when quarks $A=B$, there is an additional
disconnected correlation  $D(t)$ to be evaluated.  This  correlation can
be written in the form  
  \be D(t)= N_f r_4 r_5 L(0) L^*(t)  \ee
where the disconnected loop
 \be  L(t) =  {\rm Tr} \, { \Gamma M^{-1} } \label{loop} \ee
 with $M^{-1}$ the quark propagator  and the sum in the trace is over
colour, Dirac and spatial indices at time $t$. 

The factor of $r_5$ arises  since the Wilson-Dirac fermion matrix $M$ is
$\gamma_5$ hermitian and hence $L$ is real/ imaginary as $\gamma_5
\Gamma = r_5 \Gamma \gamma_5$ with $r_5=\pm 1$. Now at  $t=0$ we have that
$L(0) L^*(0) > 0$ so the disconnected correlation  $D(0)$ has sign $r_4
r_5$. If this sign were to be maintained at larger $t$, then this would
give a prediction for  the sign of $D/C$ and hence information on the
sign of $m_1-m_0$  without any lattice evaluation at all. The
no-free-lunch is that indeed  a sign  change in $D$ as t increases is
possible and it is indeed the goal of this work  to explore this on the
lattice.

 From Table~1, we can deduce that there must be such sign changes in
$D(t)$ as  $t$ increases from 0, since there are two operators available
to  study both pseudoscalar and vector mesons. In each case these two
operators  have different signs of $r_4 r_5$ and hence one of them must
change sign so that they agree  on a common value of $m_0 -m_1$ at large
$t$ where the ground state contribution must dominate. This has already
been explored for  the pseudoscalar case~\cite{cmcmp} and there D(t) for
$\Gamma=i \gamma_4 \gamma_5$  was found to change sign.

 \begin{table} [h]
 \begin{center}
\begin{tabular}{cccrrccc}
$\Gamma$  & $J^{PC}$ & meson& $r_4$ &  $r_5$ & $D(0)/C(t)$&
    $D(t)/D(0)$& $m_1-m_0$\\
$\gamma_5  $&$   0^{-+}  $&$\eta$,$\eta'$& -1  & 1 & -& + &-\\ 
$ i \gamma_4 \gamma_5 $&$   0^{-+}  $&$\eta$,$\eta'$& -1  & -1& + & -& - \\ 
$  \gamma_k  $&$   1^{--}  $&$\omega$,$\phi$& -1  & -1 & + & - & -\\ 
$ i \gamma_4 \gamma_k  $&$   1^{--}  $&$\omega$,$\phi$& -1  & 1 & - & + & - \\ 
$  I  $&$   0^{++}  $&$f_0$& 1  & 1 & + & + & +\\ 
$  \gamma_4  $&$   0^{+-}  $&$\hat{f_0}$& 1  & -1 & - & ? & ?  \\ 
$ i \gamma_5 \gamma_k  $&$   1^{++}  $&$h_1$& 1  & -1 & - & + & - \\ 
$ i \gamma_4 \gamma_5 \gamma_k  $&$   1^{+-}  $&$f_1$& 1  & 1 & + & - & -\\ 
\end{tabular}
 \caption{ Flavour singlet mesons produced by different operators
$\bar{\psi} \Gamma \psi$ and the sign factors $r_4$ and $r_5$ as defined
in the text whose product  determines $D(0)/C(t)$.  Here  $k$ is a
spatial index and both $\Gamma$ and $\gamma$ are hermitian. The meson 
quantum numbers are quoted in the continuum limit, for a discussion  of
the appropriate representation of the cubic group on a lattice see
\protect{\cite{hybprd}}. We also include our observed evidence for sign
changes in  $D(t)$ from $t=0$ which gives a prediction of the sign of
the non-singlet minus singlet mass shift as shown.
 }
 \end{center}
 \end{table}

\section{Lattice Methodology}

Here we use dynamical fermion configurations with  $N_f=2$ from
UKQCD~\cite{ukqcd}. The sea quarks correspond to $\kappa=0.1395$ with a 
tadpole improved clover formalism with  $C_{SW}=1.76$.  

 Local and spatially-fuzzed operators~\cite{fuzz} are used for meson
creation  (with two fuzzed links in a spatially symmetric orientation
with 5 iterations of fuzzing with  coefficient given by 2.5*Straight +
Sum of staples).  Thus we evaluate a   $2 \times 2$ matrix of local and
fuzzed correlators~\cite{fuzz}.  Mesons created by all independent
products of gamma matrices are evaluated. Here we restrict our attention
to the momentum zero sector. 

 We measure the disconnected correlations on 252 configurations of size
$12^3 24$  separated by 20 trajectories.  For the evaluation of the
disconnected correlators, we use  stochastic noise sources  with
variance reduction  using the hopping parameter expansion~\cite{cmcm}.
Here we employ more  noise sources (96) to explore the very small
disconnected contributions  from some quantum numbers. We use sources at
every site on the lattice and determine the  momentum zero correlations
from them.

 The connected correlator is obtained by explicit inversion from a
source (local or fuzzed)  at the origin for 126 configurations separated
by 40 trajectories~\cite{ukqcd}.

\begin{figure}[th]

\epsfxsize=11cm\epsfbox{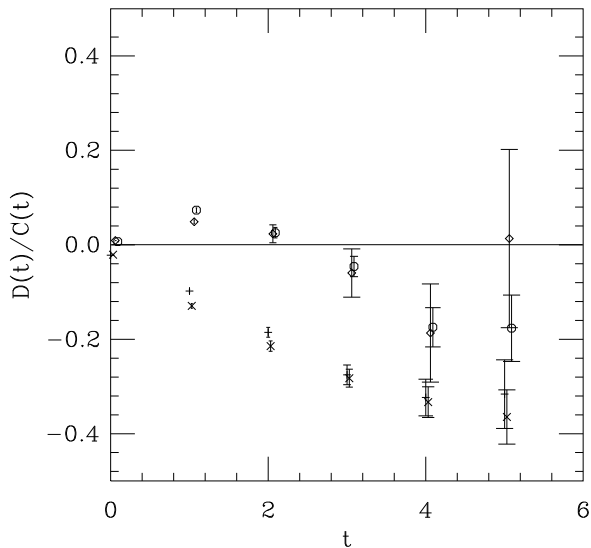}

 \caption{ The ratio of disconnected to connected contributions  for
pseudoscalar mesons versus time in lattice units. Local and  fuzzed
operators with $\Gamma=\gamma_5$\ (+, $\times$) and with $\Gamma=i
\gamma_4 \gamma_5$ (diamond, octagon).
 }
 \label{dbycp.fg}

\end{figure}

\subsection{Stochastic  noise compared to signal}

 We measure the zero momentum disconnected loop $L(t)$ on each
time-slice for  each gauge configuration. This ensemble, for each choice
of operator $\Gamma$  gives us the values of the standard deviation
$\sigma_{\rm obs}$ given in Table~2. We also,  from our 96 stochastic
samples in each case, have the estimate of the  standard deviation
$\sigma_{\rm stoch}$ on the mean of these 96 samples coming from the
stochastic method. We can then deduce the true standard deviation of the
gauge  time slices from $\sigma_{\rm gauge}=(\sigma^2_{\rm obs}-
\sigma^2_{\rm stoch})^{1/2}$. This is presented in Table~2.  Here the
normalisation is such that $M=1+\kappa \dots$.

 \begin{table} [h]
 \begin{center}
\begin{tabular}{ccccc}
$\Gamma$  & $J^{PC}$ &  $\sigma_{\rm obs}$ &  $\sigma_{\rm stoch}$ 
   &$\sigma_{\rm gauge}$ \\
$\gamma_5  $&$   0^{-+}  $&               20.82    &   5.85&19.88  \\ 
$ i \gamma_4 \gamma_5 $&$   0^{-+}  $&     9.63   &    4.05& 8.74  \\ 
$  \gamma_k  $&$   1^{--}  $           &   4.33   &    3.91& 1.86  \\ 
$ i \gamma_4 \gamma_k  $&$   1^{--}  $&    8.74   &    3.82& 7.86  \\ 
$  I  $&$   0^{++}  $&                    47.70    &   4.50&47.49  \\ 
$  \gamma_4  $&$   0^{+-}  $&              3.90   &    3.75& 1.07 \\ 
$ i \gamma_5 \gamma_k  $&$   1^{++}  $&    7.26   &    3.97& 6.08  \\ 
$ i \gamma_4 \gamma_5 \gamma_k $&$1^{+-}$&13.04    &   3.86&12.46  \\ 

\end{tabular}
 \caption{ Mesons produced by different operators $\bar{\psi} \Gamma
\psi$. The standard deviation of the loop operator of eq.~\ref{loop}  is
presented. Here $\sigma_{\rm stoch}$ is the error estimated from the  96
stochastic samples used and this is the used to deconvolute the
observed spread to give the true  standard deviation of the loop
($\sigma_{\rm gauge}$). 
 }
 \end{center}
 \end{table}

 In an ideal world we would have $\sigma_{\rm stoch} << \sigma_{\rm
gauge}$  which would imply that no appreciable error arose from the
stochastic methods  employed. For the previously studied cases, the
pseudoscalar with $\Gamma=\gamma_5$ and the scalar with  $\Gamma=I$, we
see that the stochastic errors are truly negligible with 96 stochastic
samples and indeed the 24 samples  used before~\cite{cmcm,cmcmp} were
adequate. For  the other cases, which have been little studied hitherto,
we see that the stochastic errors  are reasonably small, except  for 
$\Gamma=\gamma_k$ or $\gamma_4$.  The latter case has spin exotic
quantum numbers and is expected to be  very poorly determined by our
methods. For the vector meson, however, one  of our goals is to explore
the singlet mass splitting. Here we see that more stochastic samples 
(over 1000) would be needed with our current method  to get the
stochastic error significantly smaller than the  inherent gauge error.
We do not have the required computational resources at present. One small
advantage, however, is that we can average over the three spin
components  of the vector which reduces errors by $1/\sqrt{3}$. Also the
second operator ($\Gamma=i\gamma_4 \gamma_k$) which creates a vector
meson  has a relatively smaller stochastic error. We also use fuzzed
sources  in order to obtain more measurements.

\begin{figure}[th]

\epsfxsize=11cm\epsfbox{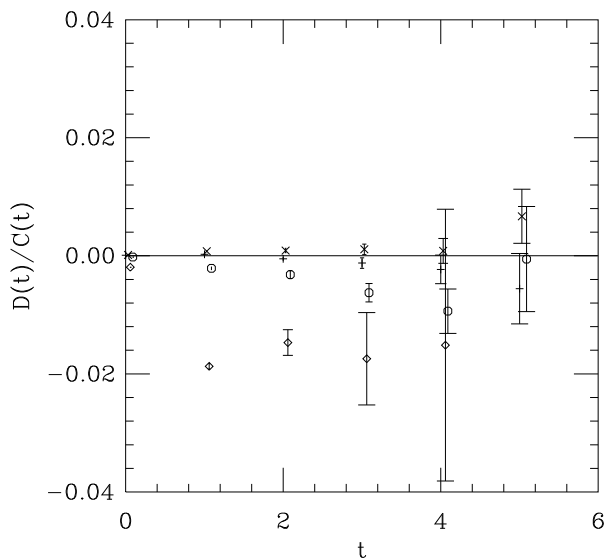}

 \caption{ The ratio of disconnected to connected contributions  for
vector mesons versus time in lattice units. Local  and fuzzed  operators
with $\Gamma=\gamma_k$\ (+, $\times$) and with  $\Gamma=i \gamma_4
\gamma_k$ (diamond, octagon).
 }
 \label{dbycv.fg}

\end{figure}

\subsection{Results}

 We present in figs.~2 to 4 some of our results for the ratio of the
disconnected  correlator to the connected correlator. The error on the 
disconnected correlator is much larger than that on the connected one. 
This arises essentially because the absolute error on the disconnected
correlator  stays of the same magnitude as $t$ increases, much as is the
case for correlations between  Wilson loops as used in glueball studies.
The connected correlator, in contrast,  has an approximately constant
relative error as $t$ increases. For this reason we employ the full data
set  to determine the disconnected correlator: sources at all sites and 
measurements every 20 trajectories. We bin these results to avoid any
problem from autocorrelations among the gauge configurations separated
by only 20 trajectories. Even with this approach, the  error on the
disconnected correlator rises rapidly as seen from the figures.  A
considerably larger number of gauge configurations will be needed to 
explore larger $t$ values.

Our results for the pseudoscalar case have been presented
before~\cite{cmcmp} and are included here for comparison. We do indeed 
see a consistent slope for the two different $\gamma$ matrix operators
considered  with a sign change for $\Gamma=i \gamma_4 \gamma_5$ as
discussed above. The mass difference $a(m_0-m_1) \approx 0.1$ is
positive here as expected  since the $\eta'$ is heavier than the
non-singlet pseudoscalar mesons.

\begin{figure}[ht]

\epsfxsize=11cm\epsfbox{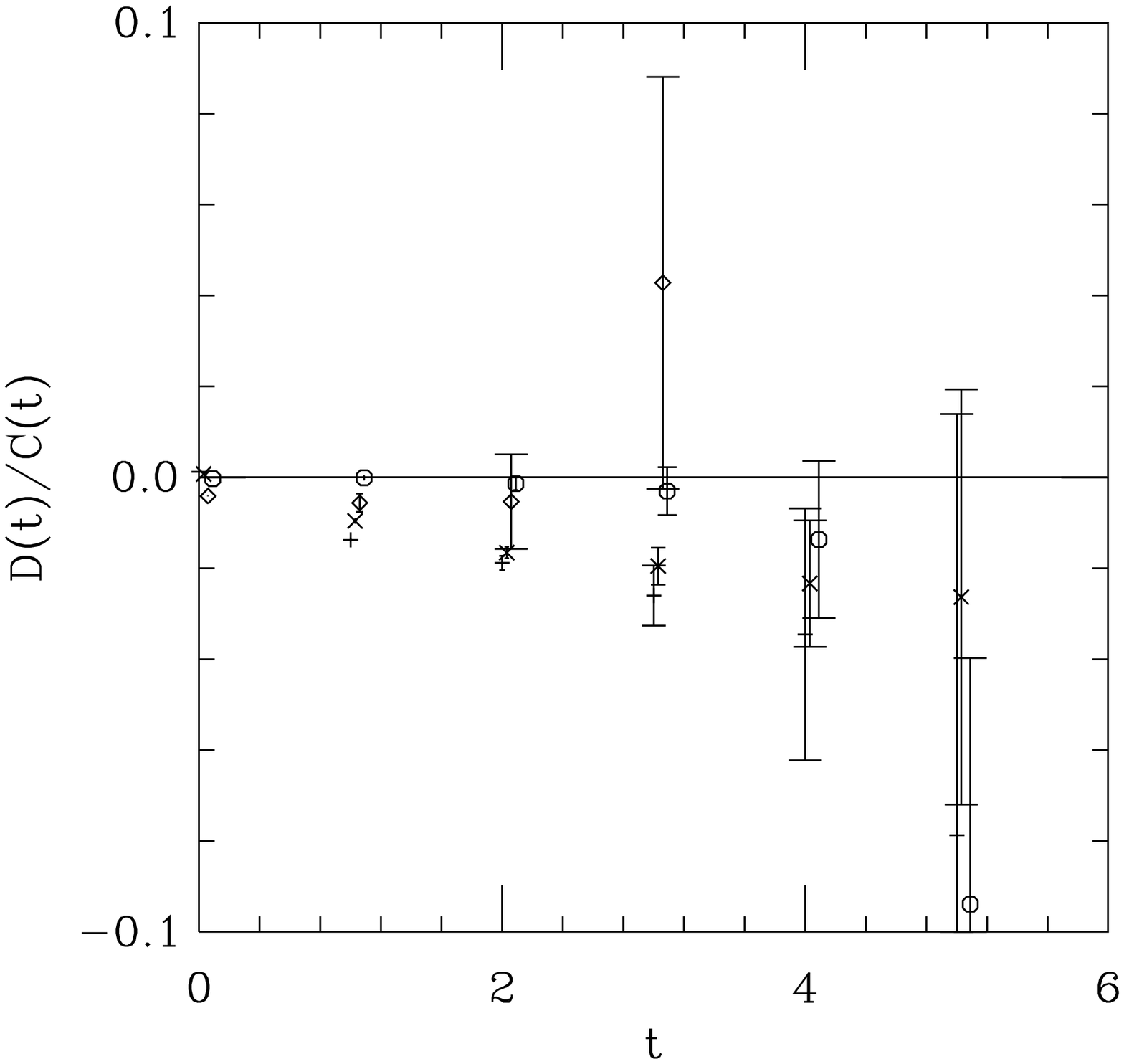}

 \caption{ The ratio of disconnected to connected contributions  for
axial mesons ($f_1$ and $h_1$ types) versus time in lattice units. Local
and fuzzed operators for the $f_1$ meson with $\Gamma=i \gamma_5
\gamma_k$\ (+, $\times$) and for the $h_1$ meson with $\Gamma=i \gamma_4
\gamma_5 \gamma_k$ (diamond, octagon).
 }
 \label{dbycab.fg}

\end{figure}

For the vector mesons, we again have two different $\gamma$ matrix
combinations available. We expect a sign change in one case as discussed
above.  From fig.~3 we see that the situation is that the ratio $D/C$ is
very  small so that detailed study is difficult. Moreover the  different
values observed for fuzzed and local operators suggest that excited
state contributions are  significant, especially for the
$\Gamma=i\gamma_4 \gamma_5$ case. The results up to  $t = 3$ do suggest
that $D/C$ is negative and that a sign change occurs for the
$\Gamma=\gamma_k$ case between $t=0$ (where it is  positive) and $t=3$.
This would imply $a(m_0-m_1) > 0$.

 For the axial mesons, our results are presented in fig.~4. The  $f_1$
meson case shows a larger signal, consistent with $a(m_0-m_1) > 0$.  We
find a sign change in $D(t)$  from $t=0$ to $t>0$ for the $h_1$
case, giving a  very small signal for $D(t)/C(t)$ at small $t$ values
which appears to be negative,  implying $a(m_0-m_1) > 0$.

 We have attempted to make fits to the singlet correlators to extract
masses in each case, but the  rapid increase in errors with increasing
$t$ makes these fits poorly determined  and with results that depend
strongly on the minimum and maximum $t$ values fitted. A less sensitive
way to estimate the allowed magnitude of the singlet  non-singlet mass
splitting is from the values of $D(t)/C(t)$ directly: bearing in mind
the  scenarios illustrated in fig.~1.  From the data shown for $t \le
5$, we estimate $a (m_0-m_1) \approx 0.10(2)$  for the pseudoscalar
case, $a (m_0-m_1) \approx 0.001(2)$ for the vector case,  $a(m_0-m_1)
\approx 0.005(10)$ for the $f_1,\ a_1$ case and $a(m_0-m_1) \approx
0.001(10)$ for the $h_1,\ b_1$ case.  Note in particular the hierarchy
of errors:  with smallest errors for the vector case, followed by the
axial mesons.  Note also, as discussed above, that we have information 
on the sign of $a (m_0-m_1)$ which is not folded into these error
estimates.

 \section{Discussion}

Our exploratory study has a scale set by~\cite{ukqcd}  $r_0/a=3.44$ and
peudoscalar meson to vector meson mass ratio of $m_{P}/m_{V} =0.71$. 
Using the conventional value $r_0=0.5$ fm then gives $a^{-1}=1.34 $ GeV
while the meson mass ratio implies that the sea quarks have masses 
close to that of a strange quark. 
 Here we are using $N_f=2$ flavours of degenerate quark in the sea. 
Since our lattice evaluation is for unphysical parameters, we first
discuss  the experimental spectrum of light quark mesons to  aid in
comparison with our results. 

 For a meson considered as made from two valence quarks of either  type
$n$ ($u$, or $d$ here treated as degenerate) or of type $s$, we  have
two observable non-singlet states $\bar{n} n$ and $\bar{n} s$. Then the 
singlet $\bar{n} n$ state would have a mass degenerate with the 
non-singlet $\bar{n} n$ if there were no disconnected contributions.
Furthermore, from the  observed non-singlet masses, by assuming an equal
 splitting in mass (or usually in mass squared), one also can deduce the mass
which the  singlet $\bar{s} s$ state would have if there were no disconnected
contributions. 
 Comparison of these  theoretical  mass values with the observed flavour
singlet masses, then gives  information from experiment about the size
and nature of disconnected contributions. 

 Thus there are  two observed states in the nonet with isospin zero which
can have a  component of the  disconnected contribution and these can mix 
in terms of their quark content. 
 So for flavour singlet sector, we then have contributions  to the mass
squared matrix with quark model content $(u \bar{u} +d
\bar{d})/\sqrt{2}$  and $s \bar{s}$ (which we label as $nn$ and $ss$
respectively):

 \be   
   \left( \begin{array}{cc} m_{nn}^2 +2x_{nn} &  \sqrt{2} x_{ns} \\
           \sqrt{2} x_{ns} &  m_{ss}^2+x_{ss}  \end{array} \right)
  \ee

 Here $m$ corresponds to the mass of the flavour non-singlet  eigenstate
as discussed above and is  the contribution to the mass coming from
connected fermion diagrams while $x$ corresponds to the contribution
from disconnected fermion diagrams. In the limit of no mixing (all
$x=0$, the OZI suppressed case), then we have the quenched QCD  result
that one  flavour singlet state is  degenerate with the isospin one
$\bar{n} n$ meson  while the other corresponds to the $s \bar{s}$ 
meson.

Using as input $m_{nn}$, $m_{ss}$ and the  flavour singlet masses  $m_0$
and $m'_0$, the three mixing  parameters $x$ cannot be fully determined.
 However if one makes some assumption about the mixing parameters $x$ 
one can deduce the mixing pattern - see \cite{cmcmp} for a discussion of
this  for pseudoscalar mesons.  One simple case that may be  used is to
assume that  all mixing strengths $x$ are the same for mesons of a 
given $J^{PC}$. Furthermore, if the value of $x^2/(m^2_{ss}-m^2_{nn})$
is small, then the off-diagonal  contribution to mixing is negligible
and the flavour singlet mass eigenstates will  be of mass squared
$m_{nn}^2+2x$ and $m_{ss}^2+x$. This gives two opportunities to estimate
$x$  from the observed spectrum. Also the mass shift of the $nn$ flavour
singlet ($2x$) is just the same as in the  case of $N_f=2$ degenerate quarks -
which is the case we explore on the  lattice here. 
 A caveat applies for axial mesons: since charge conjugation is not a
good  quantum number for the $\bar{s} n$ states, there will be mixing
between  the $J^{PC}=1^{++}$ and $1^{+-}$ mesons. This complicates the
mixing scheme  further. 

We now present the conclusions of such an analysis based on the
experimental~\cite{pdg} mass values. For the vector mesons, the $n
\bar{n}$  sector gives $m_0-m_1=0.013$ GeV while the $s \bar{s}$ sector
gives 0.016 GeV with a mass squared formalism and 0.001 GeV with a
linear mass formalism. These signs suggest that the splitting in the
$N_f=2$ sector would be  $m_0-m_1 \approx 0.01$ GeV.

 For the axial mesons, the additional mixing of the $n \bar{s}$ states
only  allows an analysis if more assumptions are made. Then assuming
that the  lightest iso-singlet state is predominantly $n \bar{n}$ yields
the two flavour result that $m_0-m_1 \approx 0.05,\ -0.06$ GeV for the
$J^{PC}=1^{++},\ 1^{+-}$ mesons  respectively.

 A complication that arises in comparing with experimental meson spectra
 is that of decays. In our lattice studies, since the quark mass is
relatively  heavy (heavier than the strange quark since  $m_P/m_V =
0.71$ while we expect~\cite{cmcmp} this ratio to be 0.682 for strange
quarks) then we have no decay channels open for the ground state mesons 
we study. In contrast, some of these mesons have substantial
experimental decay widths (150 MeV for the $\rho$, over 250 MeV for the
$a_1$ and 140 MeV for the $b_1$). One consequence of this, as has been
known for  a very long time~\cite{cmnstar}, is that the pole in the
complex plane corresponding to  a resonance has an energy whose real
part is lower than the quoted value  which corresponds to a phase shift
of $90^0$. This mass shift arises from the energy dependence of the
width  and will be more significant for wider resonances. Aside from
this inherent  uncertainty, there may be further dynamical effects
arising from the back-reaction of the decay  channels to the effective
propagator.   

 Thus we should interpret our results as giving an indication of the
strength and  sign of OZI violating contributions to the light meson
spectrum. These need not correspond to those observed experimentally 
because of the above issues (namely the more complex mixing schemes 
allowed for $N_f=3$ and the decay effects) and also because we would
need to extrapolate our lattice results to the continuum limit and to
more realistic  quark masses. In particular, there is evidence from the
pseudoscalar  mesons that the splitting does increase with decreasing
quarks mass~\cite{kuramashi,cppacs,MQA,cmcmp,sesam}. 

 We find a hierarchy of singlet non-singlet mass splitting which is large
 for pseudoscalar mesons (0.13(2) GeV), smaller for axial mesons
(0.007(13) GeV for $f_1$ and 0.001(13) GeV for $h_1$) and smallest for
vector mesons (0.002(3) GeV).  This is in agreement with the hierarchy
of magnitudes seen experimentally. We find that in each case the sign of
the effect is that the flavour singlet state is heavier. This is the
sign found experimentally for  the pseudoscalar, vector and $1^{++}$
axial but not for the $1^{+-}$ axial. The magnitudes we find are smaller
than the experimental  values (except for the $h_1$ case where the
magnitude is comparable), which can come in part from the use of too
heavy a light quark (as is  known to be the case for the pseudoscalar
mesons) but also from the impact of the large experimental decay widths.

 In this study we have explored from first principles in QCD the OZI
rule for the meson masses  for  the  case of two degenerate quark
flavours. It would be interesting to extend this lattice study to OZI
rule effects in meson decays.   We find that the disconnected
contributions (ie OZI violating terms) to the masses are indeed  small
for axial and vector mesons. 
 For the vector and axial mesons,  we find evidence that the flavour
singlet mass is increased  compared to the non-singlet. This is the
opposite of  what would be expected in the simplest theoretical model:
namely mixing with a heavier glueball of the same $J^{PC}$. An
understanding of this remains a theoretical challenge.

\end{document}